\documentclass[aip, preprint, amsmath, onecolumn ]{revtex4-2}

\usepackage{graphicx}
\usepackage{float}
\usepackage{dcolumn}
\usepackage{bm}
\usepackage{units}
\usepackage{amsmath}
\usepackage{amssymb}
\usepackage{listings}     
\usepackage[version=4]{mhchem}
\usepackage{tabularx}
\usepackage{soul}
\usepackage[normalem]{ulem}
\usepackage{array}
\usepackage{color}
\usepackage[dvipsnames]{xcolor}  
\usepackage{bookmark}
\usepackage{moresize}
\usepackage{amssymb}
\newcolumntype{P}[2]{>{\centering\arraybackslash}p{#1}}

\definecolor{dkgreen}{rgb}{0,0.6,0}
\definecolor{gray}{rgb}{0.5,0.5,0.5}
\definecolor{mauve}{rgb}{0.58,0,0.82}

\newcommand{\be}{\begin{equation}}
\newcommand{\ee}{\end{equation}}
\newcommand{\bea}{\begin{eqnarray}}
\newcommand{\eea}{\end{eqnarray}}
\newcommand{\balg}{\begin{align}}
\newcommand{\ealg}{\end{align}}

\begin{document}

\newcolumntype{L}[1]{>{\raggedright\arraybackslash}p{#1}}
\newcolumntype{C}[1]{>{\centering\arraybackslash}p{#1}}
\newcolumntype{R}[1]{>{\raggedleft\arraybackslash}p{#1}}

\author{Long Zhang}
\affiliation{Department of Physics, University of Florida, Gainesville, FL 32611, USA}
\affiliation{Quantum Theory Project, University of Florida, Gainesville, FL 32611, USA}
\affiliation{Center for Molecular Magnetic Quantum Materials, University of Florida, Gainesville, FL 32611, USA}

\author{Anton Kozhevnikov}
\affiliation{Swiss National Supercomputing Centre, Zurich, Switzerland}

\author{Thomas Schulthess}
\affiliation{Swiss National Supercomputing Centre, Zurich, Switzerland}

\author{S. B. Trickey}
\affiliation{Department of Physics, University of Florida, Gainesville, FL 32611, USA}
\affiliation{Quantum Theory Project, University of Florida, Gainesville, FL 32611, USA}
\affiliation{Center for Molecular Magnetic Quantum Materials, University of Florida, Gainesville, FL 32611, USA}

\author{Hai-Ping Cheng}
\email{hping@ufl.edu}
\affiliation{Department of Physics, University of Florida, Gainesville, FL 32611, USA}
\affiliation{Quantum Theory Project, University of Florida, Gainesville, FL 32611, USA}
\affiliation{Center for Molecular Magnetic Quantum Materials, University of Florida, Gainesville, FL 32611, USA}

\title{All-Electron APW+${lo}$ calculation of magnetic molecules with the SIRIUS domain-specific package}

\date{Dec. 2, 2022 }


\begin{abstract}

We report APW+$lo$ (augmented plane wave plus local orbital) density
functional theory (DFT) calculations of large molecular systems using the
domain specific SIRIUS multi-functional DFT package. 
The APW and FLAPW (full potential linearized APW) 
task and data parallelism options and 
advanced eigen-system solver provided by SIRIUS can
be exploited for performance gains in ground state Kohn-Sham
calculations on large systems. This approach is distinct from
our prior use of SIRIUS as a library
backend to another APW+$lo$ or FLAPW code. 
We benchmark the code and demonstrate performance on several magnetic 
molecule and metal organic framework systems. We show that the
SIRIUS package in itself is capable of handling systems as large as
a several hundred atoms
in the unit cell without having to make technical choices
that result in loss of accuracy with respect to that needed for
study of magnetic systems. 
\end{abstract}
\maketitle

\section{Introduction}

Molecular magnetism, notably in the context of molecular magnetic
materials, is a very active interdisciplinary research area that has
significant implications for computational investigations. Motivated
in no small measure by the promise of high impact on quantum computing
\cite{WasielewskiEtAl2020,GaitaArinaEtAl2019}, molecular magnetism
deals with design, synthesis, and physical and chemical
characterization of single-molecule magnets (SMMs)
\cite{doi:10.1021/ja9732439,B811963E}, spin-crossover molecules
\cite{Kepp2013,CireraViaNadalRuiz2018,Kepp2019,MntaaMejiaRodriguezEtAl}
and condensed aggregates
thereof.  Predictive and interpretive first-principles calculations
are valuable both for experimental progress and for formulation and
parametrization of models of the molecular spin systems. We begin
therefore with a brief overview of the physical problem class, then
turn to attendant computational challenges and some progress in
meeting them.

\subsection{Motivation from physical systems} 
Molecular magnetism involves a large range of dimensionality, 
from isolated SMMs \cite{MARROWS200970,annurev_conmatphys_070909_104053} and 1D chain magnets \cite{doi:10.1021/ic101016t,cyanide_chain_mag} through
2D molecular layers \cite{Zheng2016} to 3D polymers and metal organic framework \cite{acs_chemrev_9b00666,B804757J}
materials that exhibit collective ordering of magnetic moments.

Growing research activity, both theoretical and experimental, has
focused on (1) low dimensional materials (motivated by their potential
application in high-density magnetic storage and nano-scale devices)
\cite{molecular_spintronics,B616352A} and (2) so-called functional
materials
\cite{molecular_spintronics,MARROWS200970,C7CS00653E,C5SC03224E}
(because of their strong response to changing external conditions).
Because molecular magnets have well-localized magnetic moments, they
provide a nearly perfect arena for investigation of intriguing
phenomena and testing models.  Distinct from bulk magnetic materials,
their quantum size effects suggest applications beyond conventional
high-density information storage. Example applications include
spintronics and qubits for quantum computing.

The molecules involved are large and rather complicated. An example
category is Mn$_{12}$ complexes \cite{doi:10.1038/nphys877}. Their
investigation dates to synthesis by Weinland and Fischer in 1921
\cite{doi:10.1002/zaac.19211200116}, yet the crystal structure was not
determined until 1980 \cite{doi:10.1107/S0567740880007893}. That
particular Mn$_{12}$ molecule is built from four Mn$_{4+}$ ($S =3/2$) and
eight Mn$_{3+}$ ($S=2$) ions coupled by oxygen atoms. 
The 
[Mn$_{12}$O$_{12}$(O$_{2}$CPh)$_{16}$(H$_{2}$O)$_{4}$] 
complex was studied beginning in 1988, 
\cite{doi:10.1021/ja00233a036} and its ground state eventually determined\cite{doi:10.1021/ja00015a057,doi:10.1021/ja00058a027} to have $S=10$.
For computational context, note that
this system has 176 atoms and 1210 electrons.

A major class of experimental effort has focused on design and
synthesis of multi-nuclear clusters containing\cite{B811963E}
Mn$^{3+}$, because the axially Jahn-Teller distorted Mn$^{3+}$ ionic
positions usually cause large magnetic anisotropy due to spin-orbital
coupling. A large number of Mn(III)-based SMMs has been reported,
including those with Mn$_{6}$ \cite{doi:10.1021/ic900992r}, Mn$_{19}$
\cite{doi:10.1021/ja800092s,doi:10.1021/ic800213w,anie_200603498},
Mn$_{25}$ \cite{doi:10.1021/ic801142p,doi:10.1021/ja0316824},
Mn$_{31}$ \cite{jacs_7b10130} and Mn$_{84}$ \cite{anie_200353352}. In
addition, many transition metal (TM) SMMs based upon anisotropic
V$^{3+}$, Fe$^{2+/3+}$, Ni$^{2+}$, and Co$^{2+}$ ions have been
synthesized; see
Refs.~\onlinecite{doi:10.1021/ja9732439,doi:10.1021/jacs.5b08962,PhysRevLett.78.4645,doi:10.1021/ja00114a012,B108894G,doi:10.1002/anie.200351753,B605459E}.
The octa-nuclear cluster
[Fe$_{8}$O$_{2}$(OH)$_{12}$](tacn)$_{6}]^{8+}$ (Fe$^{8}$), where (tacn) is the organic compound C$_{6}$H$_{12}$(NH)$_{3}$, 
reported by Sangregorio et al.\cite{PhysRevLett.78.4645} is an example.
It has 174
atoms and 756 electrons with a $S=10$ ground state that arises from
competing anti-ferromagnetic interactions among eight Fe$^{3+}$
($S=5/2$) ions \cite{Barra_1996}. Magnetic measurements revealed that
it has an anisotropy energy barrier of 
$17 {cm^{-1}}$ induced by magnetic anisotropy.  
This is much smaller than that of the Mn$_{12}$ complexes. 
Other significant synthesis effort 
has been focused on enhancing the magnetic anisotropy of the molecule
\cite{doi:10.1021/ic701365t,C005256F,doi:10.1002/chem.200401100,B714715E}. That
has stimulated the development of several different groups of SMMs,
including cyano-bridged SMMs
\cite{doi:10.1021/ar040158e,C0CS00188K,C4CC00339J}, Ln-based SMMs
\cite{doi:10.1021/cr400018q,doi:10.1021/acs.chemrev.9b00103,SESSOLI20092328,MCADAMS2017216,ZHU2019350},
3\textit{d}-4\textit{f}-element-based SMMs
\cite{SESSOLI20092328,doi:10.1002/asia.201900897}, actinide-based SMMs
\cite{MCADAMS2017216}, radical-bridged SMMs
\cite{radical_bridged_SMMs} and organo-metallic SMMs
\cite{doi:10.1021/om401107f}.

The critical setting regarding insightful computational studies
provided by this brief overview is straightforward.  SMMs are large,
structurally and electronically intricate molecules with complicated
spin manifolds.  Materials comprised of them are, of necessity, more
complicated and demanding. Many of the chemical details of SMMs are
largely irrelevant to that assessment.  However, the presence of heavy
nuclei and the importance of anisotropy both suggest the
significance of relativistic effects, including spin-orbit
coupling. Predictive, materials-specific simulations of systems
composed from SMMs thus are extremely challenging.  Spin crossover
systems pose quite similar challenges.  

\subsection{Predictive computational approaches}

At present, density functional theory (DFT) \cite{PhysRev.136.B864} is
the most widely used first-principles calculation method in material
science.  It is based on the Hohenberg-Kohn theorems that show 
that the ground-state observables of an interacting many-electron
system are functionals of the ground-state charge density, and
that the total energy functional is variational with respect to
the density \cite{PhysRev.136.B864}.  As is well known, the HK
theorems do not provide a procedure for constructing the variational
functional.  That was provided by  Kohn and Sham who introduced
an auxiliary system of non-interacting fermions that also obey
the HK theorems \cite{PhysRev.140.A1133}. The
non-interacting system has a one-body Schr\"odinger equation
that delivers the ground state charge density, hence enables
evaluation of the variational energy.  

With an appropriate basis set, one can expand the
Kohn-Sham (KS) orbitals, potential $v_{KS}$, and charge density  and
convert the KS equations to a set of linear equations that can
be solved using refined, efficient eigen-system methods. Since $v_{KS}$ is
a density functional, the KS equations must be solved by self-consistent
iterative methods \cite{EngelDreizlerBook}. 

\begin{figure}[h]
  \centering
  \includegraphics[width=0.4\columnwidth]{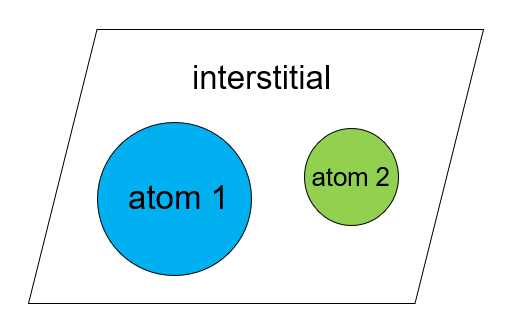}
  \caption{Muffin-tin partitioning of a unit cell.} 
  \label{fig:MT}
\end{figure}

The all-electron full-potential linearized augmented plane wave
(FLAPW) basis is one well-tested form for implementation of the KS
problem. A related form is APW+$lo$, augmented plane waves plus
local orbitals.  This family of basis sets  is based on a partitioning
of the  unit cell of a material 
into non-overlapping muffin-tin (MT) spheres, centered at the atomic
nuclei, and an interstitial (IS) region between the MT spheres, as
illustrated in Fig.~\ref{fig:MT}.

These basis sets originated from the APW method proposed by Slater
\cite{PhysRev.51.846,SLATER196435}. It involved a non-linear eigenvalue
problem for the determination of the MT basis functions themselves.
A significant step forward was the introduction 
of linearized methods 
by Andersen \cite{PhysRevB.12.3060}and applied by Koelling
and Arbman \cite{Koelling_1975} using the muffin-tin
approximation to the potential. They also pointed out that the FLAPW
made non-muffin-tin calculations feasible.
The basis
set is constructed according to the same space partition. One piece of
the basis function for Bloch wave-vector $\bm{k}$ and plane-wave
vector $\bm{G}$ is 
\be 
\varphi_{\bm{k}}^{\bm{G}}(\bm{r}) =
\begin{cases}
\displaystyle
\sum_{\ell,m} \sum_{\nu} A_{\ell m \nu}^{\alpha,\bm{k}}(\bm{G}) u_{\ell \nu}^{\alpha}(r) Y_{{\ell}m}(\hat{\bm{r}}) , & {\bm r} \in \alpha \\
\displaystyle (1/\sqrt{\Omega })  e^{i({\bm G}+{\bm k})\cdot {\bm r}} , & {\bm r} \notin \alpha 
\end{cases} 
\label{origAPW}
\ee
Here $u_{l\nu}^{\alpha}(r)$ is the solution of the (energy dependent)
radial Schrodinger equation in the MT sphere labeled $\alpha$,
$Y_{\ell m}(\hat{\bm r})$ are spherical harmonics,
$A_{\ell m\nu}^{\alpha,\bm{k}}(\bm{G})$ are the matching coefficients for
connection with the interstitial plane wave, $\ell$ and $m$ are the
azimuthal and magnetic quantum numbers in a particular sphere, and
$\nu$ is the order of energy derivative of the radial function.  The
APW basis set does not have continuous radial first derivatives at the
sphere boundaries.  The dependence of the radial functions upon the
energy for which they are solved is the APW basis difficulty.
Continuity at sphere boundaries requires those energies
to correspond to KS eigenvalues.  That correspondence makes the KS
secular equation highly non-linear in the one-electron energies. The
problem is remedied by introducing the energy derivative of the radial
function to the basis, ${\dot u}_{\ell\nu}^{\alpha} = [\partial /
  \partial \epsilon]u_{\ell\nu}^{\alpha}$. The resulting 
linearized APW (LAPW) basis is ordinary, in the sense of being
decoupled from explicit dependence on the KS eigenvalues.
The basis can be enhanced by addition of
local orbitals ($lo$), which are radial functions and energy
derivatives that vanish at the MT boundaries.

For any of these APW-based forms, the electron density and 
the effective potential are expanded in correspondence to the
space division in the unit cell.  In the interstitial region they are
expanded in plane waves, while inside the MT spheres the expansion is
in real spherical harmonics $R_{\ell m}({\bm r})$: 
\be n({\bm r}) =
\begin{cases}
\displaystyle
\sum_{\ell m} n_{\ell m}^{\alpha}(r) R_{\ell m}(\hat {\bm r}) , & \bm{r} \in \alpha  \\
\displaystyle \sum_{\bm{G}} \tilde{n}(\bm{G}) e^{i{\bm G}\cdot{\bm r}},  & \bm{r} \notin \alpha
\end{cases} 
\label{densityexpansions}
\ee
and 
\be
v_{KS}({\bm r})=
\begin{cases} \displaystyle
\sum_{\ell m} v_{\ell m}^{\alpha}(r) R_{\ell m}(\hat {\bm r}) , & \bm{r} \in \alpha  \\
\displaystyle \sum_{{\bm G}} \tilde{v}({\bm G})  e^{i{\bm G}\cdot{\bm r}} ,  & \bm{r} \notin \alpha \; .
\end{cases}
\label{vksexpansions}
\ee
Here $n_{\ell m}^{\alpha}(r)$, $\tilde{n}({\bm G})$,
$v_{\ell  m}^{\alpha}(r)$, and $\tilde{v}({\bm G})$ are expansion
coefficients determined through the self-consistent solution of the KS
equation.

Notice that the muffin-tin decomposition of the unit cell is used solely for
constructing the basis functions and expanding various quantities. The
KS equation that is solved has no shape approximation, hence the
phrase ``full-potential''.  There also is no reliance on
pseudo-potentials or projector augmented waves.  Thus FLAPW and
APW+$lo$ solutions are all-electron treatments.  For that reason, they
 commonly are considered and used as a highly precise (``gold
 standard'') realization of DFT against which the precision of other
 computational implementations of DFT is tested.

Precisely because the methods using APW-type basis sets are full
potential and all-electron, there is a serious issue for using them as
gold-standard tests in large-scale computational investigations of
molecular magnetic materials.  That key issue is efficient use of
machine resources and achievement of meaningful computational speed.
Addressing that issue motivates the focus of the present work upon the
domain-specific SIRIUS package.  The design goal of SIRIUS was to
achieve computational efficiency.  Our goal here is to exploit SIRIUS 
to make fast and efficient all-electron, full-potential DFT
calculations routinely feasible for large and complicated systems as
exemplified by magnetic molecules and their aggregates. In the
following sections, first we describe the package and its intended
usage for accelerating plane-wave based DFT codes.  We then explain
its value when used as a stand-alone FLAPW/APW+$lo$ package. Then we
demonstrate its capabilities by calculations on a selection of
magnetic molecules.

\section{The SIRIUS package}

Irrespective of the particular code, FLAPW/APW+$lo$ calculations
obviously have the same underlying formalism. 
Because those basis sets start from plane waves,
such codes also have significant elements in common with plane-wave
pseudo-potential (PW-PP) codes. Those elements include:
unit cell setup, atomic configurations, definition and
generation of reciprocal lattice vectors $\bm G$ and
combinations with Bloch wave vectors ${\bm G} + {\bm k}$, 
definition of basis functions on regular grids as Fourier expansion
coefficients; construction of the plane-wave contributions to
the KS Hamiltonian matrix, generation of the charge density, effective potential, and magnetization on a regular grid;
symmetrization operations on the charge density, potential and occupation matrix; 
iteration-to-iteration mixing schemes for density and potential;
diagonalization of the secular equation.  As already is evident, compared to PW-PP
codes, FLAPW/APW+$lo$ codes also have everything expanded in radial functions
and spherical harmonics inside the MT spheres, along with enforcement
of matching conditions on sphere surfaces.

As we have discussed elsewhere \cite{computation10030043}, those commonalities
offer an opportunity, namely \textit{separation of concerns}. 
For those generally common tasks in FLAPW/APW+$lo$ and PW-PP codes,
the concept is to create a performance-optimized package with enough
user support that it can be exploited either as a basic code or as a
library.  This strategy allows the
abstraction and encapsulation of the common objects just listed,
thereby exposing opportunities for optimizing computational
performance irrespective of the user interface (input,
post-processing, etc.). The SIRIUS package was designed from the
outset and developed from the outset with that goal. It provides both
task and data parallelization.  It is optimized for multiple MPI
levels as well as OpenMP parallelization and for GPU utilization as
well.

SIRIUS can be interfaced directly with both existing FLAPW/APW+$lo$
codes and with PW-PP codes as a \textit{DFT library}.  We have
presented an example of such use elsewhere
\cite{computation10030043}. Though that was the originally intention
for the primary usage of SIRIUS, 
it turns out to have some limitations
imposed by design incompatibilities structured into the host
code. However, SIRIUS does have basic FLAPW/LAPW+$lo$ stand-alone
capability. Thus there is opportunity to see what benefits can be
gained for large-scale magnetic system calculations in that mode of
SIRIUS usage.

A few details about SIRIUS are relevant. 
It is written in C++ along with the CUDA \cite{cudaToolKit} backend
to provide several features: (1)
low-level support such as pointer arithmetic and type casting as well
as high-level abstractions such as classes and template
meta-programming; (2) easy interoperability between C++ and widely
used Fortran90; (3) full support from the standard template library
(STL) \cite{cpp_stl_lib}; (4) easy integration with the CUDA nvcc compiler \cite{cuda_nvcc_compiler}. Though not relevant here, SIRIUS provides
dedicated API functions for interfacing to 
the QuantumEspresso and \textit{exciting} codes.  

Motivated by the demands of large-system calculations, SIRIUS is
designed and implemented with both task distribution and data (large
array) distribution in mind.  Note that typical KS calculations in a
basis set rely on two basic functionalities, distributed complex
matrix-matrix multiplication ({\tt pzgemm} in LAPACK
\cite{LAPACK_129995}) and a distributed generalized eigenvalue solver
({\tt pzhegvx} also in LAPACK). SIRIUS handles these two major tasks
with data distribution and multiple levels of task distribution.

For computational capability, switching from LAPACK to ScaLAPACK gives
the benefit of data parallelism but does not remove the
diagonalization algorithm limitation.  Switching from LAPACK (or
ScaLAPACK) to a Davidson-type diagonalization addresses that.  Doing
so requires taking into consideration both the diagonalization
algorithm and the handling of large data sets as the system size
grows. This is especially so in the case of iterative diagonalization
implemented in FLAPW/LAPW+$lo$ basis codes. When the unit cell is as
large as ten \AA$^3$ and contains a hundred or more multi-electron
atoms, the plane-wave cutoff needed is normally about 25--30
$a_0^{-1}$ (where $a_0$ is the Bohr radius)  to reach a properly
converged ground state. This causes the reciprocal-lattice-vector
($G$-vector) related arrays to become very large.

The problem is illustrated by the ELK and \textit{exciting}
codes. Each has several multi-dimensional arrays that have one
dimension for the global $G$-vector indices. Because those arrays are
not handled in a distributed way, they become very memory consuming in
a single MPI task.  Eventually they become the real bottleneck, once
the diagonalization algorithm limitation is removed via a
Davidson-type method. SIRIUS, in contrast, treats the $G$-vector
related multi-dimensional arrays in distributed manner by design.

As noted already, standard libraries do not offer Davidson-type 
diagonalization algorithms because they utilize repeated
application of the Hamiltonian to a sub-space of the system.
Algorithmic implementation therefore depends on construction details
of the Hamiltonian matrix, i.e. it depends on the specific DFT
formalism. Again, the SIRUS package takes this into consideration
by design. It provides an efficient implementation of Davidson-type 
diagonalization \cite{Gulans_Davidson_Iterative_solver} for both 
FLAPW/APW+$lo$ and PW-PP codes. 

With the foregoing discussion for context, we turn to examples of
stand-alone SIRIUS calculations on large molecular magnetic and
spin crossover systems that exemplify our interests.


\section{SIRIUS: [Mn(taa)] molecule} 

Because it has proven to be a difficult case
\cite{MntaaMejiaRodriguezEtAl}, the first system studied is a
molecular spin crossover (SCO) complex called [Mn(taa)].  Such systems often are found
in transition metal compounds and metal organic frameworks.  The SCO
transition can be induced by a variety of external perturbations, e.g.
as temperature, pressure, magnetic, and electric field.  SCO molecular
materials have potential utility for reversible molecular switching
in functional materials.  The [Mn(taa)] molecule ([$Mn^{3+}
  (pyrol)_{3} (tren)$])  is a meridional
pseudo-octahedral chelate complex of a single Mn ion as the magnetic
center and the hexadentate
tris[(E)-1-(2-azolyl)-2-azabut-1-en-4-yl]amine ligand.  It has 53 atoms and
224 electrons.  
Originally studied by Sim and Sinn \cite{doi:10.1021/ja00391a067}, it was 
the first known example of a Mn(III) 3$d^{4}$ SCO system.
Experimentally it is an example
of a transition that can be induced by application of
an external magnetic
field. Without any external magnetic field,
the Mn$^{3+}$ cation goes from a low-spin (LS) state (S=1) to a
high-spin (HS) state of (S=2) at a transition temperature of about
$45 \, \textrm{K} $.  The total energy difference between LS and HS
states, $\Delta E_{HL} = E^{tot}_{HS} - E^{tot}_{LS}$, is relatively
 small, which makes magnetic-field-induced SCO transition
experimentally feasible.  Since the SCO is caused by the change of
3$d$ electron configuration, the transition to the HS state is
accompanied with not only a variation of dielectric constant but also
an enhancement of the lattice volume.  An important feature of the SCO
transition of [Mn(taa)] is that the reorientation of the electric
dipole moments appears to arise from the dynamic Jahn-Teller (JT)
effect in the HS state \cite{10.1063/1.5097891}.  Detailed 
first-principles calculation in combination with Monte Carlo simulation
disclosed three competing HS phases: a Jahn-Teller
ordered (solid) phase, a dynamically correlated (liquid) phase, and an
uncorrelated (gas) phase \cite{PhysRevLett.124.227201}.  The system
poses challenges to the computational determination of the ground
state due to the aforementioned small scale of $\Delta E_{HL}$ relative
to the total energy magnitudes.  Estimates
are about $ 50 \pm 30 \, \textrm{meV} $ but as high as a few hundred
meV depending upon the exchange-correlation (XC) approximation used
and various technical details \cite{MntaaMejiaRodriguezEtAl}.  Note
that usual XC approximations suffer from a self-interaction error and
hence tend to favor the LS state to reduce spurious self-repulsion
with the result of overestimated $\Delta E_{HL}$ values.  That is not
a concern in the current work since what we are testing is the
capability of SIRIUS in finding the LS ground state.

Several factors can affect a DFT calculation of the 
molecular  ground state  significantly. For consistency with condensed phase
calculations, it is appropriate to study the isolated molecule in a
large, periodically bounded box.
To attain required accuracy, the plane wave cutoff needs to be large.
For the isolated molecule, the vacuum volume in the computational
unit cell requires even larger cutoffs.  
This makes the APW+$lo$ calculation an intensive job because of the
comparatively large 
Hamiltonian matrix to be diagonalized and the large $G$-vector related arrays. 

The quality of APW+$lo$ calculations is governed by the dimensionless
quantity $R^{MT}_{min} \cdot |\bm{G}^{k}|_{max}$, where $R^{MT}_{min}$
is the minimum muffin-tin radius of all atomic species and
$|\bm{G}^{k}|_{max}$ is the maximum $\bm{G}+\bm{k}$
vector magnitude. A common default value of $R^{MT}_{min} \cdot |\bm{G}^{k}|_{max}$
is 7 in most FLAPW/APW+$lo$ codes. It can be 10 if one wants to push the
accuracy of total energy to be close to $\mu$Ha. For organic
molecules, the smallest $R^{MT}$ is normally around $ 1.4 \, a_{0}$ for 
an H atom, which puts $|\bm{G}^{k}|_{max} \in [5, 7] \,a^{-1}_{0} $. The size
of the Hamiltonian matrix in the first variational
step is determined by $|\bm{G}^{k}|_{max}$, the total number of atoms
and the local orbitals added to each atom.

For [Mn(taa)], if one sets $R^{MT}_{min} \cdot |\bm{G}^{k}|_{max}=7$
and other parameters as shown in Table \ref{tbl:MntaaLS}, the
dimension of the first variational Hamiltonian matrix is about
330,000. This is actually within the theoretical limit of the full
diagonalization algorithm in LAPACK. However, in order to run the DFT
calculation, not only the $\bm{G}^{k}$ vectors but also the $\bm{G}$
vectors for expanding density and potential must be handled
efficiently. For molecules in a large cubic unit cell of the size of
20 \AA{} with irregular vacuum space, we found (from SIRIUS
calculations) that the $|\bm{G}|$ for density and potential needs to
be as large as 25-- $30 \, a^{-1}_{0}$ to make the total energy
converge. If the arrays containing $\bm{G}$ vector indices are not
stored in a distributed way, they become the major memory consuming
quantities, which makes the calculation very slow.

The difficulty is real.  We have tried to use community codes such as
ELK and Exciting-Plus to calculate the ground state of [Mn(taa)] and
found it to be quite difficult. If one performs a single $k$-point
calculation for an isolated system, the lack of band parallelism
within a $k$-point makes the run have only one MPI task. As just
noted, the large $|\bm{G}|$ cutoff needed for density and potential
implicates significant memory allocation that makes the calculation
slow. A test to run the [Mn(taa)] system using Exciting-Plus on the
NERSC Cori system (128GB memory per node) with $|\bm{G}| = 30 \,
a^{-1}_{0}$ or larger caused a crash out-of-memory (OOM) error, even
if one node was assigned with only one MPI task.

To have a more specific measure of the memory issue we ran the same
test on the Univ. Florida HiperGator HPC large memory node and
observed the peak memory consumption values reported in
Table-\ref{tab:EP_memory}.


For the SIRIUS calculation of [Mn(taa)], we used the experimentally
determined low spin structures and the PBE
\cite{PhysRevLett.77.3865}  generalized gradient approximation
(GGA) XC functional. We set $R^{MT}_{min} \cdot |\bm{G}^{k}|_{max}=5$,
$|\bm{G}|_{max}=30 \, a^{-1}_{0}$, and other input parameter values
listed in Table-\ref{tbl:MntaaLS}.  There was no Hubbard $U$. The same
structures and XC were used in a VASP \cite{vasp} calculation for
comparing the results; see Table-\ref{tbl:MntaaLS}. 

The ground state HOMO-LUMO gap from SIRIUS is 3\% larger than that
from the VASP calculation. The difference is small so that it does not
suggest significant difference in physical or chemical properties.
The total magnetization for the two calculations agrees at
2.00 $\mu_{B}$. The Mn atom moments are 1.68 $\mu_{B}$ and
1.77 $\mu_{B}$ from SIRIUS and VASP respectively. The difference arises
from the fact that the muffin-tin radius for Mn in SIRIUS is 2.2
$a_{0}$, compared to the smaller PAW projector radius of for Mn
in VASP, which is 2.8 $a_{0}$. We defer comparative discussion of
average time per SCF iteration to Section \ref{timing}. 

The benefit of this particular study
is to show that SIRIUS can be used in ``gold-standard'' mode in
studies of large molecular complexes, hence
 to confirm, in particular, that our previous VASP calculations on the [Mn(taa)]
  complex were sufficiently accurate for the purpose. 

\begin{table}
  \centering
  \begin{tabular}{ | C{4cm} | C{2.75cm} | C{2.75cm} | C{2.75cm} | C{2.75cm} | } 
  \hline
  $|\bm{G}|_{max}$ ($a^{-1}_{0}$) & 20    & 25     & 30     & 35      \\  \hline
  peak memory (GB)                & 39.5  & 110.8  & 151.8  & 209.5   \\  \hline
  \end{tabular}
  \caption{Peak memory consumption of single $k$-point [Mn(taa)] calculation in a cubic unit cell of $ 20\,$\AA. Other input parameters were the same as Table.\ref{tbl:MntaaLS}.} 
  \label{tab:EP_memory}
\end{table}

\begin{figure}[h]
  \centering
  \includegraphics[width=0.4\columnwidth]{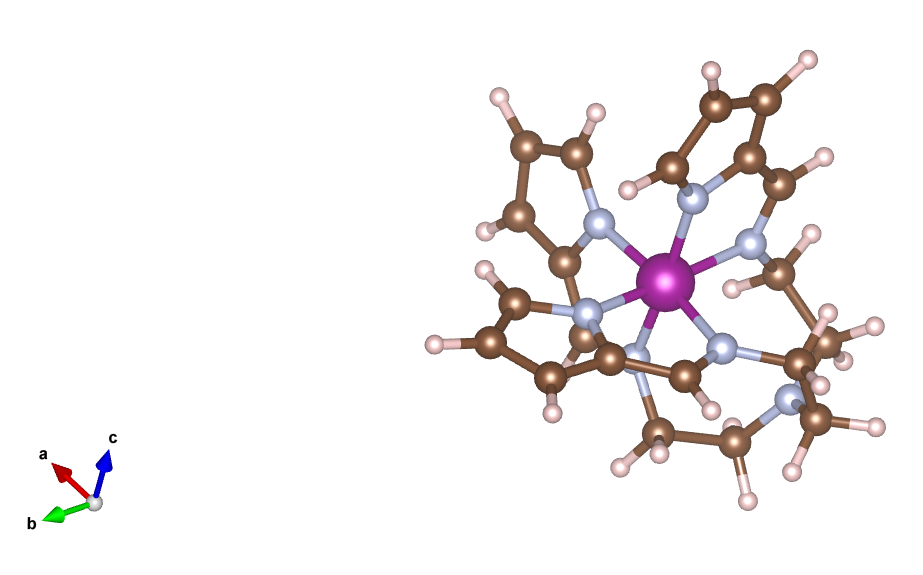}
  \caption {[Mn(taa)] molecule}  
  \label{fig:31taa}
\end{figure}

\begin{center}
\begin{table}[h]
\caption{Input parameters for the [Mn(taa)] LS state
\label{tbl:MntaaLS}}
\begin{tabular}{ | c | c | } 
\hline
\multicolumn{1}{|l|}{structure}                                  & Mn-taa, LS state structure \\ \hline
\multicolumn{1}{|l|}{unit cell}                                  & $20\times20\times20$ \AA{}  box   \\ \hline
\multicolumn{1}{|l|}{number of atoms}                            & 53     \\ \hline
\multicolumn{1}{|l|}{$R_{mt}$ ($a_{0}$)}                         & Mn: 2.2; O: 1.6; C: 1.4; H: 1.0;  \\ \hline  
\multicolumn{1}{|l|}{$R^{MT}_{min} \cdot |\bm{G}^{k}|_{max}$}    &  5     \\ \hline   
\multicolumn{1}{|l|}{$|\bm{G}_{max}|$ ($a_{0}^{-1}$) for $\rho$ and $V_\textrm{eff}$}      & 30     \\ \hline
\multicolumn{1}{|l|}{$l_{max}$ for APW}                                  & 8      \\ \hline  
\multicolumn{1}{|l|}{$l_\textrm{max}$ for $\rho$ and $V_\textrm{eff}$}   & 8      \\ \hline
\multicolumn{1}{|l|}{size of 1st variational Hamiltonian}   &  $\approx$ 155,000      \\ \hline
\multicolumn{1}{|l|}{$k$-point grid }                           & $1\times1\times1$     \\ \hline
\multicolumn{1}{|l|}{(L)APW configuration }                      & $\epsilon_{l}=-0.15 \,$eV; $\partial_{E}=0$;  \\
\multicolumn{1}{|l|}{$lo$ configuration }                        & Mn: $s,p,d$; O/C: $s,p$; H: $s$; \\ \hline
\multicolumn{1}{|l|}{treated as core state}                      & Mn: 1s, 2s, 2p, 3s; O/C: 1s              \\ \hline

\multicolumn{1}{|l|}{total energy tolerance}                     & $10^{-6}$ Har              \\ 
\multicolumn{1}{|l|}{potential tolerance}                        & $10^{-7}$ Har              \\  \hline
\multicolumn{1}{|l|}{run job setup:}                    & 16 MPI tasks             \\ 
\multicolumn{1}{|l|}{             }                     & 16 OMP threads per task  \\
\multicolumn{1}{|l|}{number of SCF iterations}          & 75                       \\
\multicolumn{1}{|l|}{average time per SCF iteration}    & 55 s                     \\  \hline

\multicolumn{1}{|l|}{HOMO-LUMO gap (eV) }                & 0.68    (VASP: 0.66)     \\ \hline
\multicolumn{1}{|l|}{$\mu_{tot}$ ($\mu_{B}$)    }        & total: 2.00  (VASP: 2.00) \\ 
\multicolumn{1}{|l|}{   }                                & Mn atom: 1.68 (VASP: 1.77)        \\ \hline  
\end{tabular} 
\end{table} 
\end{center} 

\section{SIRIUS: $Mn_{3}$ dimer molecule} 

One highly desirable property to demonstrate for potential application
of SMMs would be quantum mechanical coupling of two or
more SMMs either to one other or to a surface or other device component, all
the while retaining their isolated-molecule magnetic properties to a
useful degree.  For this, an SMM-SMM coupled structure was identified
for hydrogen-bonded supramolecular pairs of $S=9/2$ SMMs
[$\textrm{Mn}_{4}\textrm{O}_{3}\textrm{Cl}_{4}(\textrm{O}_{2}\rm{CEt})_{3}(\textrm{py})_{3}$]
\cite{Hill1015,doi:10.1063/1.2355106,PhysRevLett.91.227203}.  Since
hydrogen-bonded inter-SMM interactions do not provide easy control of
oligomerization nor guarantee retention of the oligomeric structure in
solution, covalently organic linked SMM-SMM structures were developed.

The first along that line was the $[\textrm{Mn}_{3}]_{4}$ SMM tetramer
\cite{doi:10.1021/ja2087344}, which is covalently linked with
dioximate linker groups. A recent further development is the
$[\textrm{Mn}_{3}]_{2}$ dimer molecule
\cite{doi:10.1021/jacs.5b02677}, comprised of two $\textrm{Mn}_{3}$
units covalently joined via dpd2-dioximate linkers. The two triangular
$\textrm{Mn}_{3}$ units are parallel and the inter-$\textrm{Mn}_{3}$
interaction has been determined to be ferromagnetic. There is no
experimental evidence for noticeable interactions between the two
$\textrm{Mn}_{3}$ units.  That is consistent with the absence of
significant inter-$\textrm{Mn}_{3}$ contacts in the structure and the
relatively large distance between Mn ions.  The interaction quantum
mechanically couples the two $\textrm{Mn}_{3}$ units. The structure is
robust and resists any significant deformation or distortion that
might affect the weak inter-$\textrm{Mn}_{3}$ coupling.

With the same consideration as for Mn[(taa)], we studied the isolated
$[\textrm{Mn}_{3}]_{2}$ SMM dimer in a large periodically bounded box
to be consistent with condensed phase calculations and to be assured
of isolation. Note that 
$[\textrm{Mn}_{3}]_{2}$ has 137 atoms and 748 electrons, which is a
much larger system then Mn[(taa)].

There is an important distinction.  Both the $[\textrm{Mn}_{3}]_{2}$ SMM
dimer and [Mn(taa)] calculations are done with a single $k$-point.
Both relied upon the band parallelization in SIRIUS.  The next two
(see below) rely on that band parallelization as well. Both of those
are metal-organic framework structures that do not have irregular
vacuum regions and can use moderate Brillouin zone scan meshes, hence
$k$-point parallelization as well.  As with the [Mn(taa)] example,
this $[\textrm{Mn}_{3}]_{2}$ SMM dimer calculation also benefits from
distributed storage of the $G$-vector related arrays.

Specifically, the molecule also was placed in a $20 \, $\AA{} periodically
bounded box with single $k$-point Brillouin Zone sampling.  The
$G_{max}$ cutoff and LO/$lo$ configuration for the Mn atoms were the
same as for [Mn(taa)]. We also used the experimental
geometry \cite{doi:10.1021/jacs.5b02677} and the PBE
XC functional \cite{PhysRevLett.77.3865} (again, no
Hubbard $U$).  We are able to converge the system to the
anti-ferromagnetic ground state in which there are two spin-up Mn and
one spin-down Mn in each $[\textrm{Mn}_{3}]$ unit, and the total moment is zero.

The SIRIUS calculation yields similar values to those from VASP in the
HOMO-LUMO gaps, which are almost identical. For the magnetization,
the VASP value for the Mn ion is larger than what SIRIUS gives.
Apparently this is traceable to the muffin-tin radius (SIRIUS)
being smaller than the projector cutoff
radius (VASP). The average time consumed per SIRIUS SCF iteration is about
2.5 times larger than that for [Mn(taa)], but this is based on the same 16
MPI tasks for band parallelization. Increasing the number of tasks or
number of cores per task would definitely make the job run faster.
Comparison with VASP timings is in the next section.  As with the
previous study, this one 
shows that the SIRIUS APW+$lo$ calculation can be used as a ``gold-standard'' 
to test and check VASP results. 

\begin{figure}[h]
  \centering
  \includegraphics[width=0.3\columnwidth]{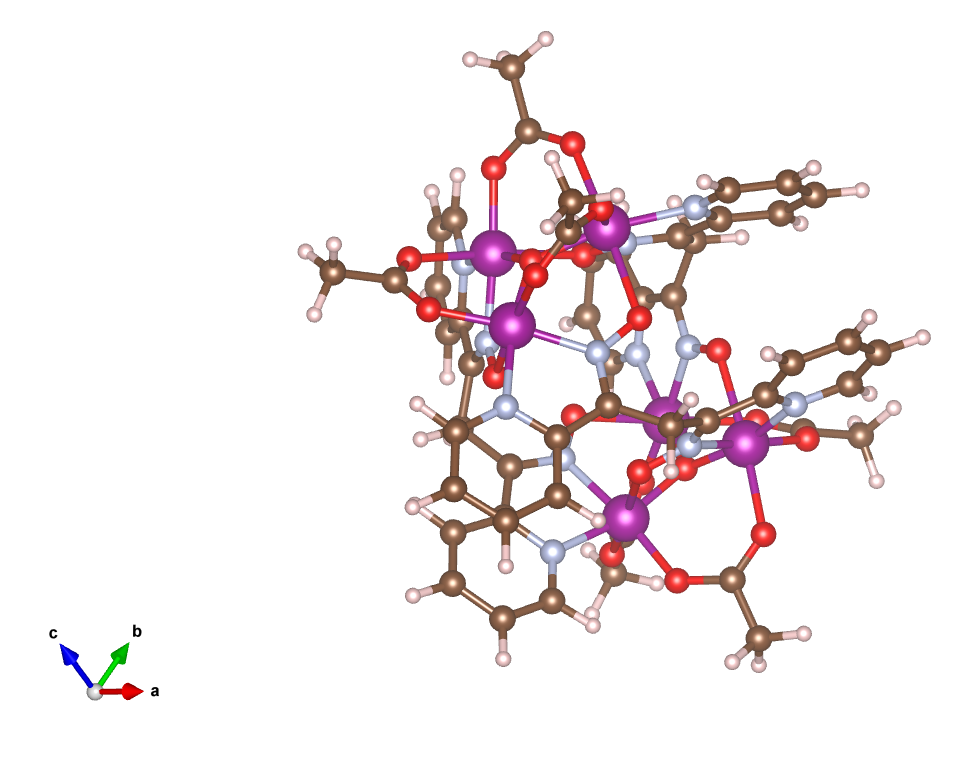}
  \caption {Mn$_3$ dimer molecule}  
  \label{fig:31dimer}
\end{figure}

\begin{center}
\begin{table}
\caption{Input parameters and outputs of $\textrm{Mn}_{3}$ dimer
\label{tbl:Mn3}}

\begin{tabular}{ | c | c | c | } 
\hline
            & $\textrm{Mn}_{3}$ dimer                    \\ \hline
\multicolumn{1}{|l|}{unit cell}                                  & $20\times20\times20 \, $\AA{} cubic                 \\ \hline
\multicolumn{1}{|l|}{number of atoms in unit cell}               & 137                              \\ \hline
\multicolumn{1}{|l|}{$R_{mt}$ ($a_{0}$)}                         & Mn: 2.2; O/N: 1.4; C: 1.2; H: 1.0; \\ \hline  
\multicolumn{1}{|l|}{$R^{MT}_{min} \cdot |\bm{G}^{k}|_{max}$}    &  5     \\ \hline   
\multicolumn{1}{|l|}{$|\bm{G}_{max}|$ ($a_{0}^{-1}$) for $\rho$ and $V_\textrm{eff}$}      & 30     \\ \hline
\multicolumn{1}{|l|}{$l_{max}$ for APW}                                  & 8      \\ \hline  
\multicolumn{1}{|l|}{$l_\textrm{max}$ for $\rho$ and $V_\textrm{eff}$}   & 8      \\ \hline
\multicolumn{1}{|l|}{size of 1st variational Hamiltonian}   &  $\approx$ 187,000      \\ \hline
\multicolumn{1}{|l|}{$k$-point grid }                           & $1\times1\times1$     \\ \hline
\multicolumn{1}{|l|}{(L)APW configuration }                      & $\epsilon_{l}=-0.15\,$eV; $\partial_{E}=0$; \\
\multicolumn{1}{|l|}{$lo$ configuration $\star$}                 & Mn: $s,p,d$; O/C/N: $s,p$; H: $s$ \\ \hline

\multicolumn{1}{|l|}{core state}             & Mn: 1s, 2s, 2p, 3s; O/C/N: 1s   \\ \hline

\multicolumn{1}{|l|}{total energy tolerance}            & $10^{-6}$ Har      \\ 
\multicolumn{1}{|l|}{potential tolerance}               & $10^{-7}$ Har  \\  \hline
\multicolumn{1}{|l|}{run job setup:}                    & 16 MPI tasks             \\ 
\multicolumn{1}{|l|}{             }                     & 16 OMP threads per task  \\
\multicolumn{1}{|l|}{number of SCF iterations}          & 125                   \\
\multicolumn{1}{|l|}{average time per SCF iteration}    & $\approx$ 250 s         \\  \hline
\multicolumn{1}{|l|}{HOMO-LUMO gap (eV) }                & 0.26    (VASP: 0.27)     \\ \hline
\multicolumn{1}{|l|}{$\mu_\textrm{tot}$ ($\mu_{B}$)    } & each Mn atom (Sirius): $\pm  1.92$ \\ 
\multicolumn{1}{|l|}{   }                                & each Mn atom (VASP): $\pm  1.95$ \\ \hline  

\end{tabular}
\end{table}
\end{center}

\section{Timing Comparison \label{timing}} 

Caution must be exercised in interpreting what follows.   This is 
because comparison of computational costs between conceptually and structurally
distinct codes is a task fraught with difficulties.  Optimal resource
assignments generally differ greatly not only between codes but among
different machine configurations. Both the definitions and
effects of precision tolerances differ between codes as well.

We compare SIRIUS only with VASP because of its very wide use and
because of the cost of running the large systems studied here with
other PW-PP codes. Even this restriction to one comparison involves
some critical choices.  Begin with MPI parallelization.  We can force
the two codes to run with the same number of MPI tasks, but the
situations for each task are very different.  The discussion that
follows is specific to the NERSC Cori machine.

By experiment, we have found that the optimum number of OpenMP threads per
task for SIRIUS is 16. To obtain adequate memory per task, it also
is necessary to assign no more than two tasks on a node. Putting
more than two on a single node degrades performance noticeably.

In contrast, for VASP the recommended practice
does not require use of OpenMP.  If, however, we use openMP for the sake
of even-handed comparison, experiment showed that ordinarily four openMP
threads per task generally is optimum (with eight at most).  Thus,
we can put four or more tasks per node.  

Our testing was for the two systems reported above, [Mn(taa)] and
Mn$_3$ dimer.  For both we did single $k$-point calculations at fixed
geometries with 16 MPI tasks and optimal OpenMP threads per task for
each code.  The choice of diagonalizer is unique in SIRIUS but not in
VASP.  In it, the better choice appears to be the tag "ALGO=Fast",
which invokes a mixture of the Davidson and RMM-DIIS algorithms.

With these choices, the results are that the ratio of average time per
SCF iteration for VASP versus SIRIUS is about 50\% for [Mn(taa)] and
about 25\% for the Mn$_3$ dimer.  For [Mn(taa)] the iteration count is
75 for SIRIUS and 55 for VASP, so the rough computational cost of
SIRIUS versus VASP is about a factor of 2.7.  For the Mn$_3$ dimer,
the iteration counts are 125 (SIRIUS) and 90 (VASP), so the
computational cost ratio is about 5.6.  

Use of the VASP ``ALGO=normal'' option, which invokes a blocked
Davidson diagonalization algorithm, increases the VASP iteration count
by about 10\%, so the ratios become roughly 2.4 and 5.1 respectively.

Of course these cost ratios refer to different numbers of electrons,
since SIRIUS is all-electron while VASP uses PAWs.  For [Mn(taa0] the
the electron count is 224 versus 156 (SIRIUS and VASP respectively)
while for the Mn$_3$ dimer, the counts are 748 and 510 respectively.
If cubic scaling with electron number (which is the most elementary
expectation with Kohn-Sham calculations) is followed, the
[Mn(taa)] ratio would be 2.96, not far from the observed 2.7  For
Mn$_3$, however, cubic scaling would give 3.16.  Apparently the
actual 5.6 is a consequence of parallelization but we have not
investigated the matter.

The larger point is that the cost-scaling, though significant, is not
prohibitive even for quite large molecular systems.  With that
established, the next two sections give two more illustrative examples
of the use of SIRUS in the context of molecular magnetic quantum
materials.


\section{SIRIUS: DTN molecule} 

\begin{figure}[h]
  \centering
  \includegraphics[width=0.75\columnwidth]{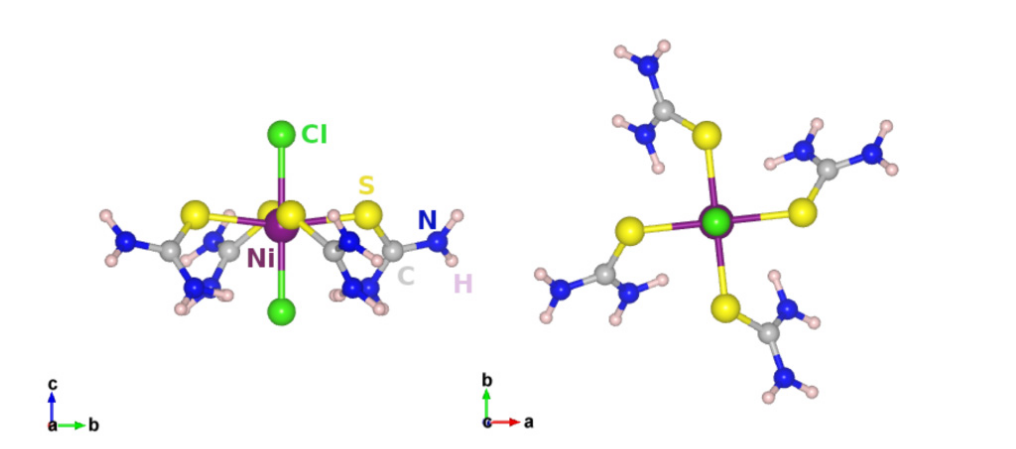}
  \caption {base octahedral unit of DTN molecule \cite{PhysRevB.103.054434}}  
  \label{fig:DTN_base}
\end{figure}

The insulating organic compound
$\textrm{XCl}_{2}$-$[\textrm{SC(NH}_{2})_{2}]_{4}$ with X=Ni or Co
(DTN or DTC molecule, respectively), is a molecule-based framework
structure in which magnetic and electric order can couple. It
has been studied experimentally for its quantum magnetism
\cite{PhysRevLett.96.077204,PhysRevLett.98.047205,PhysRevB.83.140405,PhysRevB.77.020404}. ($\textrm{NiCl}_{2}$-$[\textrm{SC(NH}_{2})_{2}]_{4}$)
has a tetragonal molecular crystal structure with two Ni atoms as
magnetic centers in one unit cell. Each Ni has four S atoms and two Cl
atoms as nearest neighbors, as shown in Fig.~\ref{fig:DTN_base}.
They form  an octahedral structure similar to the $\textrm{BO}_{6}$
octahedra in $\textrm{ABO}_{3}$ perovskites. The base octahedral
structures pack in a body-centered tetragonal lattice, with
Ni–Cl–Cl–Ni bonding along the $c$-axis (Fig.~\ref{fig:DTN_base}) and
hydrogen bonding in the $a$-$b$-plane. The four thiourea
$[\textrm{SC(NH}_{2})_{2}]$ groups around each Ni ion are electrically
polar. The $a$-$b$-plane components of their electric polarization
cancel, while the $c$-axis components are in the same direction,
thereby creating a net $c$-axis electric polarization that could be
responsible for magnetic field-modified ferroelectricity.

At temperatures below $ 1.2 \, \textrm{K} $ and below a critical
magnetic field, DTN is a quantum paramagnet \cite{PhysRevB.103.054434}. 
As the magnetic field along the $c$-axis reaches the first critical value,
DTN undergoes a quantum phase transition into an
XY-antiferromagnetic state in which all Ni spins lie within the
$a$-$b$-plane. Upon further increase of the field, the spins begin to
be more aligned with a corresponding increase in magnetization. When a
second critical field is reached, the magnetization saturates and the
material enters a spin-polarized state with all spins aligned parallel 
to the applied magnetic field.

There are three non-equivalent Ni-Ni spin couplings in the DTN bulk
that are of interest: $J_{c}$ along the $c$-axis (along the
Ni–Cl–Cl–Ni bonding) between two Ni ions in adjacent unit cells,
$J_{ab}$ in the $a$-$b$-plane between two Ni ions in adjacent unit
cells, and $J_{diag}$ between the two Ni ions within one unit cell.
Experiments suggest that the exchange coupling along the $c$-axis via
the Ni-Cl-Cl-Ni chain is strong \cite{nature_dtn}, and the system
behaves as a quasi one-dimensional AFM chain of Ni$^{2+}$ ions such
that each ion has $S=1$.  DFT calculation can capture the main
differences in values between the exchange coupling constants.  A
recent DFT+$U$ study \cite{PhysRevB.103.054434} showed that rather
large Hubbard $U$ correction (about 5--$6\,$eV) is needed to match
experimentally fit values of the coupling constants.  However the main
feature, namely that $J_{c}$ is about an order of magnitude larger than
$J_{ab}$, is already captured by ordinary KS-DFT calculations without
the Hubbard $U$ correction.

Thus, we calculated  $J_{ab}$ and $J_{c}$ using SIRIUS without $U$.
To estimate the $J$ values, we first calculate the total energy of the
FM ordered and the AFM ordered states, $E_{FM}$ and $E_{AFM}$,
respectively.  Then, assuming a Heisenberg Hamiltonian of the form
$H=J [S_{1} \cdot S_{2}]$, one can determine $J$ from
$E_{FM}$-$E_{AFM}$ \cite{doi:10.1063/1.5127956}.  The calculation is
done using supercells, $1 \times 1 \times 2$ for $J_{c}$ and $2 \times
1 \times 1$ for $J_{ab}$.  The $J_{diag}$ is obtained with the
primitive unit cell.  The supercells contain a number of atoms similar
to that for the Mn$_{3}$-dimer molecule in the preceding example,
namely 140 atoms and 888 electrons. The size of the supercell is
approximately $10 \times 10 \times 20 \, $\AA$^{3}$, about
half of the cell used for both [Mn(taa)] and the Mn$_{3}$ dimer.  Though the
unit cell is smaller, we found proper convergence in total energy
still requires a high plane-wave cutoff, $|G_{max}|$=25-30
($a_{0}^{-1}$), for density and potential.  The input parameters are
listed in Table-\ref{tbl:DTN}.

Using the APW+$lo$ basis and with quality parameter $R^{MT}_{min}
\cdot |\bm{G}^{k}|_{max}=6$, we can get the converged PM ground state
using the primitive unit cell and the AFM and FM ground states using
1x1x2 and 2x1x1 supercells. In the AFM and FM configurations, the calculated
magnetic moment of Ni is slightly smaller than the VASP 
value.   The
$J_{ab}$ and $J_{c}$ determined from the differences in total
energies, Table-\ref{tab:DTN_J}, confirmed the observation obtained
from using VASP. The ratio $J_{ab}/J_{c} \approx 8$ is actually
slightly more significant than that determined by VASP, and both are
in reasonable agreement with experiments. When using the primitive unit cell to
determine $J_{diag}$, we found little differences in total energies of
the FM and AFM ground state. This does not really conflict with the
VASP observed 0.03 meV because the limit of the FLAPW accuracy
(excluding approximations brought by XC functionals) is at 1 $\mu$Har
which is exactly 0.027 meV. The VASP-observed very small $J_{diag}$
could be within the PAW pseudo-potential approximation error bar.  


\begin{center}[H]
\begin{table}
\caption{Input parameters and outputs of DTN \label{tbl:DTN}}

\begin{tabular}{ | c | c | c | } 
\hline
        & DTN $1 \times 1 \times 2$ supercell      \\ \hline
\multicolumn{1}{|l|}{unit cell}                                  & $\approx$ $10\times10\times20\,$\AA{} box  \\ \hline
\multicolumn{1}{|l|}{number of atoms in unit cell}               & 140      \\ \hline
\multicolumn{1}{|l|}{$R_{mt}$ ($a_{0}$)}                         & Na/Co: 2.2; Cl/S/N: 1.4; C: 1.2; H: 1.2;  \\ \hline  

\multicolumn{1}{|l|}{$R^{MT}_{min} \cdot |\bm{G}^{k}|_{max}$}    &  6     \\ \hline   
\multicolumn{1}{|l|}{$|\bm{G}_{max}|$ ($a_{0}^{-1}$) for $\rho$ and $V_\textrm{eff}$}      & 30     \\ \hline
\multicolumn{1}{|l|}{$l_\textrm{max}$ for APW}                                  & 8      \\ \hline  
\multicolumn{1}{|l|}{$l_\textrm{max}$ for $\rho$ and $V_\textrm{eff}$}   & 8      \\ \hline
\multicolumn{1}{|l|}{size of 1st variational Hamiltonian}   &  $\approx$ 225,000      \\ \hline

\multicolumn{1}{|l|}{$k$-point grid}                            & $ 2 \times 2 \times 2 $    \\ \hline
\multicolumn{1}{|l|}{(L)APW configuration }                      & $\epsilon_{l}=-0.15$ eV; $\partial_{E}=0$; \\
\multicolumn{1}{|l|}{for $l \leq l_\textrm{max}^\textrm{APW}$}   &    \\ \hline

\multicolumn{1}{|l|}{$lo$ configuration}                 & Na/Co: $s,p,d$  \\ 
\multicolumn{1}{|l|}{  }                                         & Cl/S/C/N: $s,p$; H: $s$   \\ \hline

\multicolumn{1}{|l|}{treated as core state}                      & Ni/Co: 1s,2s,2p,3s           \\
\multicolumn{1}{|l|}{  }                                         & Cl/S/C/N: 1s              \\ \hline

\multicolumn{1}{|l|}{total energy tolerance}                     & $10^{-7}$ Har              \\ 
\multicolumn{1}{|l|}{potential tolerance}                        & $10^{-7}$ Har              \\  \hline
\multicolumn{1}{|l|}{run job setup:}                    & 64 MPI tasks             \\ 
\multicolumn{1}{|l|}{             }                     & 16 OMP threads per task  \\
\multicolumn{1}{|l|}{number of SCF iterations}          & 90                       \\
\multicolumn{1}{|l|}{average time per SCF iteration}    & $\approx$ 450 s                     \\  \hline
\multicolumn{1}{|l|}{HOMO-LUMO gap (eV) }                & 0.18    (VASP: 0.18)     \\ \hline
\multicolumn{1}{|l|}{magnetic moment }       & Ni atom: $\pm 0.89$ (VASP: $\pm 0.92$)        \\ \hline   

\end{tabular}
\end{table}
\end{center}

\clearpage

\begin{table}[h]
  \centering
  \begin{tabular}{ | C{4cm} | C{3cm} | C{3cm} | C{3cm} | } 
  \hline
                & $J_{c}$ (meV) & $J_{ab}$ (meV) & $J_{diag}$ (meV) \\  \hline
  SIRIUS (DTN)  & $-2.72$         & $-0.34$          & 0.00            \\  \hline
  VASP (DTN) \cite{PhysRevB.103.054434} & $-1.05$         & $-0.15$          & 0.03            \\  
  \hline
  \end{tabular}
  \caption{Exchange coupling constants from Sirius calculation compared to VASP results. \cite{PhysRevB.103.054434}} 
  \label{tab:DTN_J}
\end{table}

\clearpage

\section{SIRIUS: [Fe($tBu_{2}qsal)_{2}$] molecule} 

The [Fe$(tBu_{2}qsal)_{2}$] molecule is a recently created spin
crossover MOF structure \cite{doi:10.1021/jacs.1c04598}, where
$(tBu_{2}qsal)$ stands for
2,4-diterbutyl-6-((quinoline-8-ylimino)methyl)phenolate.  The
crystallized structure was determined by X-ray diffraction to be
monoclinic in space group $P2_{1}/c$. It is potentially a functional
molecular magnetic material because [Fe$(tBu_{2}qsal)_{2}$] can be
sublimed at 473--$573 \, \textrm{K} $ and $10^{-3}$--$10^{-4}\,$mbar,
hence its thin-film deposition on a substrate is possible.  The system
undergoes a hysteretic spin transition.  The average Fe–N and Fe-O
bond lengths elongate from 1.949(2) \AA{} at $ 100 \, \textrm{K} $ to
2.167(2) \AA{} at $230\, \textrm{K}$ and from 1.945(1) at $ 100 \,
\textrm{K} $ to 1.997(1) \AA{} at $230\, \textrm{K}$, respectively.
These changes indicate that conversion from the LS ($S=0$) to HS
($S=2$) structure takes place as the temperature is increased.  In
addition, PW-PP-based DFT calculations \cite{Le_2021} found the
electron transfer is minimal for the molecule on a monolayer of
Au(111). It suggests very small changes in the electronic structure
and magnetic properties when the molecule is placed on the surface of
Au.

For a system of this size, 436 atoms, 2435 electrons, we used
$R_{min}^{MT} \cdot G_{max} =4$ and other cutoffs as listed in
\ref{tbl:Fe4}.  The large number of atoms does not make the
Hamiltonian matrix dimension significantly larger than in the
preceding cases but it makes the calculation much more time consuming.
This is because the total number of radial functions of the APW
functions linearly increases with the number of atoms.  For each
muffin-tin the radial functions are normally expanded up to $l=8$, in
comparison with the $\beta$ projectors
\cite{PhysRevB.41.7892,PhysRevB.59.1758} of the pseudo-potential method
which typically are defined for $s$, $p$ and $d$ states only.  In each
SCF iteration, diagonalization of the Hamiltonian involves repeated
application of the Hamiltonian to a set of wave functions.  That
requires summation over the APW radial functions, a task that becomes
a major time consuming operation when treating large systems using
APW+$lo$.  For this MOF structure, we used the experimental structure
and single $k$-point Brillouin zone sampling. The $R^{MT}_{min} \cdot
|\bm{G}^{k}|_{max}$ was set to 4.  The plane wave cutoffs for density and
potential were set to 30, as before, and the angular momentum cut-off
was set to the normal value,  8.  Maximum benefit of band
parallelization on the NERSC CORI machine
(which has two 16-core CPUs and total 128GB memory per node)
is reached at 8x8 MPI tasks.
Each task is assigned with the resources of 16 cores thus two tasks
occupy one node, which is tested to be optimal. Further increasing the
number of tasks, or assigning more resources to each task, doe not
seem to accelerate the calculation.  With the setup described, we
managed to converge the system to non-magnetic ground state in about
80 hours. The total time needed to converge to the
  ground state also depends on the number of SCF iterations, which
  depends on the mixing algorithm for the charge density. The mixing
  algorithm used in the present work is the Broyden scheme
  \cite{Broyden1965} in which the Jacobian matrix is approximated from
  the previous step and improved iteratively. Other mixing schemes
  that implicitly approximate the inverse Jacobian from multiple
  previous steps, e.g. the Anderson mixing scheme
  \cite{AndersonMixing}, are being tested.  However adjusting and testing the mixing parameter or
any other input parameter using a system of this size is not
really practical, simply as a matter of resource availability.  Thus the
calculation of this MOF structure is a feasibility test only.  We
have shown meaningful KS-DFT calculations \textit{can be done} on a  system of
this size using APW+$lo$ and have confirmed the computational
cost advantage of efficient job distribution.






\begin{figure}[h]
  \centering
  \includegraphics[width=0.75\columnwidth]{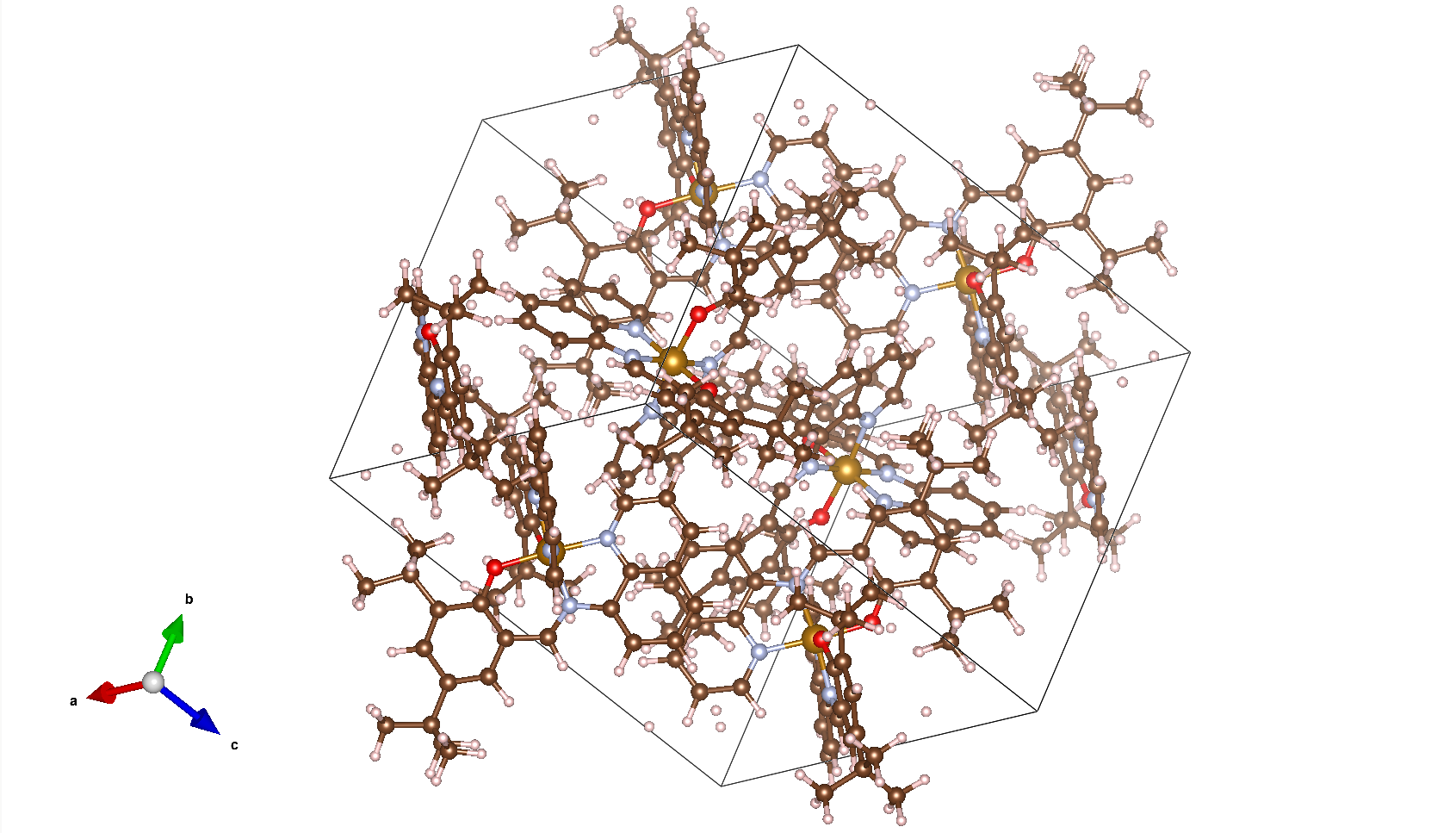}
  \caption {the Fe4 molecule}  
  \label{fig:Fe4molecule}
\end{figure}

\begin{center}
\begin{table}
\caption{Input parameters of $\textrm{Fe}_{4}$ molecule, S=0
\label{tbl:Fe4}}

\begin{tabular}{ | c | c | c | } 
\hline
                                                                 & $\textrm{Fe}_{4}$ MOF                    \\ \hline
\multicolumn{1}{|l|}{unit cell}                                  & $15\times16\times17.5 \,$\AA                 \\ \hline
\multicolumn{1}{|l|}{number of atoms in unit cell}               & 436                              \\ \hline
\multicolumn{1}{|l|}{$R_{mt}$ ($a_{0}$)}                         & Fe: 2.2; O: 1.4; C: 1.2; H: 1.0; \\ \hline  
\multicolumn{1}{|l|}{$R^{MT}_{min} \cdot |\bm{G}^{k}|_\textrm{max}$}    &  4     \\ \hline   
\multicolumn{1}{|l|}{$|\bm{G}_{max}|$ ($a_{0}^{-1}$) for $\rho$ and $V_\textrm{eff}$}      & 30     \\ \hline
\multicolumn{1}{|l|}{$l_\textrm{max}$ for APW}                                  & 8      \\ \hline  
\multicolumn{1}{|l|}{$l_\textrm{max}$ for $\rho$ and $V_\textrm{eff}$}   & 8      \\ \hline
\multicolumn{1}{|l|}{size of 1st variational Hamiltonian}   &  $\approx$ 325,000      \\ \hline

\multicolumn{1}{|l|}{$k$-point grid}                            & $1\times1\times1$                                     \\ \hline
\multicolumn{1}{|l|}{(L)APW configuration }                      & $\epsilon_{l}=-0.15\,$eV; $\partial_{E}=0$;        \\ \hline

\multicolumn{1}{|l|}{$lo$ configuration $\star$}                 & Fe: $s,p,d$               \\ 
\multicolumn{1}{|l|}{  }                                         & Cl/S/C/N: $s,p$; H: $s$   \\ \hline
\multicolumn{1}{|l|}{treated as core state}                      & Ni/Co: 1s, 2s, 2p, 3s   \\
\multicolumn{1}{|l|}{  }                                         & Cl/S/C/N: 1s  \\ \hline

\multicolumn{1}{|l|}{total energy tolerance}                     & $10^{-6}$ Har  \\ 
\multicolumn{1}{|l|}{potential tolerance}                        & $10^{-7}$ Har  \\  \hline
\multicolumn{1}{|l|}{run job setup:}                    & 64 MPI tasks             \\ 
\multicolumn{1}{|l|}{             }                     & 16 cores per task  \\
\multicolumn{1}{|l|}{             }                     & 16 OMP threads per task  \\
\multicolumn{1}{|l|}{number of SCF iterations}          & 135       \\
\multicolumn{1}{|l|}{average time per SCF iteration}    & $\approx$ 2150 s       \\ \hline

\end{tabular}
\end{table}
\end{center}

\clearpage
\section{Summary and Conclusion} 

The main purpose of this work is to demonstrate, by several concrete
examples, the value of the SIRIUS architecture and implementation for
all-electron KS-DFT calculations on fairly large molecules with
complicated spin manifolds.  By inference, this capacity extends to
aggregates of such systems.  SIRIUS as a stand-alone package
provides performance gains through refined diagonalization methods and task
and data parallelization improvements. The result is an advance in
capability of the APW+$lo$ DFT treatment of complex molecular aggregates. 

To recapitulate, the eigenvalue solver and the distribution of $G$-vector
arrays in community FLAPW codes, for example ELK, \textit{exciting}
and Exciting-Plus, are major bottleneck in calculations of large
systems. The LAPACK/ScaLAPACK full diagonalization algorithm cannot
handle Hamiltonian matrices larger than $\approx 10^{6}$, and the
$G$-vector arrays cannot be handled efficiently without distributed
storage. The SIRIUS package provides better data distribution and
options to the use of the various (LAPACK, ScaLAPACK, Davidson)
diagonalization algorithms. One can perform Davidson-type
diagonalization of the Hamiltonian easily in the self-consistent loop and
benefit from multiple MPI and thread-level parallelization within
$k$-points and within bands. Together with the proper distribution of
$G$-vector related arrays, the SIRIUS package can do plane wave based
LAPW and APW+$lo$ calculation of systems larger than many community
LAPW/APW+$lo$ codes.

We showed results from molecules using the APW+$lo$ basis. The
resulting total energy and magnetization are in agreement with
experimental measurements and confirm corresponding VASP calculations
done with PAWs. In
these test calculations, good scaling in band parallelization is
observed, which is particularly crucial for a single $k$-point
calculations ([Mn(taa)], Mn$_{3}$ dimer). The results also indicate
that SIRIUS parallelization works well on contemporary high
performance systems and the computational time is drastically reduced
compared to ELK and Exciting-Plus for example.
For a system of the size of DTN supercell studied in this work, the main physical feature
is captured by extraction of exchange constants $J$
from the total energies. The results are qualitatively in agreement
with experiment and VASP calculations. 



Looking ahead, all-electron FLAPW and APW+$lo$ calculations of medium to 
large molecule and MOF systems is a relatively little-explored area.
It is plausible that important core effects from, for example, spin-orbit
coupling will be uncovered by such all-electron investigations.  At the
least, the use of plane-wave-PAW codes to drive ab initio BOMD will
be validated at sample configurations by such all-electron calculations.
We have shown that 
the SIRIUS package can handle systems as large as 200 non-H atoms routinely
without losing the accuracy needed for magnetic systems.  With suitable
high-performance systems, SIRIUS demonstrably  can be used for
systems up to 430 atoms and, we surmise, larger.  

~

\noindent \textbf{Acknowledgment} This work was supported as part of
the Center for Molecular Magnetic Quantum Materials, an Energy
Frontier Research Center funded by the U.S. Department of Energy,
Office of Science, Basic Energy Sciences under Award No. DE-SC0019330.
Computations were performed at NERSC and the University of Florida
Research Computer Center.

\clearpage

\bibliography{main.bib}

\end{document}